\documentclass[submission, Phys]{SciPost}
\usepackage{graphicx}
\usepackage{amsmath}
\usepackage{amsfonts}
\usepackage{amssymb}
\usepackage{epsfig,libertine}
\usepackage[libertine]{newtxmath}
\usepackage{color}
\usepackage{dsfont, hyperref, xcolor}
\usepackage{comment}
\usepackage{enumerate}
\usepackage{mathtools}
\usepackage{physics}
\usepackage{array, dcolumn}
\begin{document}

\begin{center}{\Large \textbf{Transport properties and thermopower of the spinful Sachdev-Ye-Kitaev dot}}\end{center}
\begin{center}
{Daniele Morotti, Marco Uguccioni, Luca Dell'Anna}
\end{center}
\begin{center}
{Dipartimento di Fisica e Astronomia e Sezione INFN, Universit\`{a} degli Studi di Padova, via Marzolo 8, 35131 Padova, Italy}
\end{center}

\date{\today}

\section*{Abstract}
{\bf We study the electric and thermoelectric transport through a spinful complex Sachdev-Ye-Kitaev (SYK) quantum dot coupled to metallic leads, forming a N-SYK-N junction, by the Keldysh field theory approach. Unlike traditional equilibrium approaches, our formulation treats the system as an open, interacting quantum conductor under non-equilibrium conditions, without resorting to the replica trick. Starting from the exact Keldysh-Dyson equations, we derive analytical results for the tunneling and zero-temperature limits and perform a numerical analysis in the linear-response regime. We characterize the dependence of conductance, thermoelectric coefficient, and Seebeck effect on the particle-hole asymmetry parameter and coupling strength to the leads. Our results reveal distinctive non-Fermi liquid signatures of the SYK model in transport properties and identify coupling regimes where thermoelectric effects are enhanced, suggesting experimentally accessible fingerprints of SYK physics in mesoscopic systems.}

\noindent\rule{\textwidth}{1pt}
\tableofcontents\thispagestyle{fancy}
\noindent\rule{\textwidth}{1pt}

\section{Introduction}
Understanding electron transport at the mesoscopic scale remains a cornerstone of modern condensed matter physics. In this intermediate regime, between the microscopic atomic scale and macroscopic conductors, quantum coherence dominates the dynamics, giving rise to interference patterns, quantized conductance, and fluctuation phenomena beyond classical intuition.

Unlike traditional solid-state physics, where materials are treated as closed and at equilibrium, mesoscopic systems are inherently open and driven: they are connected to external electrodes or reservoirs, and their dynamics are governed by continuous exchanges of particles and energy. This places the problem firmly in the realm of non-equilibrium quantum physics, whose complete understanding is considered one of the most challenging tasks in modern theoretical physics. However, the mesoscopic regime is not just a playground for theory; it underpins quantum technologies such as charge sensing, quantum thermoelectricity, superconducting qubits, and nanoscale heat engines \cite{bib:1new,bib:2new,bib:3new,bib:4new}. In other words, a comprehensive theoretical understanding of mesoscopic systems is essential for both fundamental physics and device engineering.

Two of the most fundamental and experimentally accessible building blocks in mesoscopic physics are the quantum point contact (QPC) \cite{bib:qpc} and the quantum dot (QD) \cite{bib:5new,bib:6new}. These systems not only serve as testbeds for understanding quantum transport at the nanoscale, but also provide practical platforms for applications in quantum technologies, from charge sensing to thermoelectric energy harvesting, qubit architectures 
{\color{black}and heat engineering \cite{bib:josefsoon,bib:benenti}. 
Thermopower and thermoelectric properties in these systems have been also investigated  in the past \cite{bib:sivan,bib:beenakker92,bib:staring,bib:dzurak}. }

Beyond the experimentally established platforms like QPCs and QDs, there has been growing interest in exploring more exotic systems where quantum coherence and strong interactions dominate. Among these, the Sachdev-Ye-Kitaev (SYK) model \cite{bib:SY,bib:kitaev} has attracted considerable attention for its remarkable properties and connections to both condensed matter and high-energy physics. 
The strong interest shown towards this model stems from a rare combination of properties. In fact, it is an exactly solvable quantum many-body system (in the large $N$ and infrared limit), based on $N$ strongly coupled disordered fermions with random all-to-all interactions, that shows maximally chaotic dynamics and an emergent approximate conformal symmetry. This 
is also a toy model for holography, in the context of high-energy physics, being connected holographically to black holes with 2D anti-de Sitter horizons \cite{bib:kitaev, bib:maldacena}. 
Other studies showed that this model and its generalizations have also relevant properties in the context of condensed matter physics, such as information scrambling, chaos, quantum entanglement and strange metallic (i.e. non-Fermi liquid) behavior \cite{bib:trunin,bib:chowdhury}. Concerning the experimental realization of the model, there are many proposals either in solid state physics \cite{bib:SYKexp1} and in cavity quantum electrodynamics platforms \cite{bib:SYKexp2}. However, to date, no definitive realization of the full SYK model has been achieved, though significant progress has been made in designing intermediate systems that capture some aspects of its behavior.

Despite this wide-ranging interest, much of the theoretical work on SYK systems has focused on equilibrium properties or thermodynamics. The problem of quantum transport through a SYK system, especially when it is connected to external metallic leads and driven out of equilibrium, remains 
less explored, although it attracted already some theoretical interest 
\cite{bib:gnezdilov,bib:can, bib:kruchkov,bib:cheipesh, bib:pavlov, bib:francica, bib:dellanna}. Transport probes provide access to fundamentally different information than static quantities: they are sensitive to the nature of excitations, relaxation dynamics, and coupling to the environment.

{\color{black} Several transport theories} \cite{bib:landauer,bib:buttiker,bib:petruccione} 
are often not suitable when addressing open, driven interacting systems where electrons exchange energy and particles continuously with reservoirs. 
A versatile, real-time, ab-initio theoretical approach is provided by the 
Keldysh Field Theory (KFT) \cite{bib:keldysh,bib:kamenev,bib:diehl}, which systematically captures both transient dynamics and non-equilibrium steady states (NESS). KFT is a functional integral approach 
which offers a transparent, systematic framework for studying interacting open systems, including QPCs, QDs, and SYK islands, naturally handling real-time dynamics, dissipation, non-equilibrium steady states, quenched disorder, and connections to Lindbladian and classical stochastic dynamics. It also bypasses limitations of linear-response or equilibrium techniques \cite{bib:fetter}, making it a versatile tool and the ideal candidate for a fundamental theory of non-equilibrium quantum physics. 
We applied already KFT to investigate transport properties of quantum dots coupled to both normal-metallic (N) and superconducting (S) leads \cite{bib:uguccioni,bib:uguccioni2},
showing that this approach 
is a powerful tool for addressing quantum transport problems. 

The aim of the present work is therefore twofold. First, we show that KFT proves to be particularly well suited for treating quenched disordered systems such as the SYK model, as it circumvents cumbersome mathematical procedures and their associated subtleties, most notably the well-known replica trick \cite{bib:edwards} that is traditionally employed in the standard SYK literature. More precisely, we  
derive the exact Keldysh-Dyson equations by employing KFT for the SYK model, 
either isolated and coupled to metallic leads. Second, as already done in the case of a N-QD-N junction \cite{bib:uguccioni}, we perform a clear and systematic study of the electric and thermoelectric currents in a spinful complex SYK model connected to metallic reservoirs, i.e. a N-SYK-N junction, 
deriving the Seeback coefficient from the explicit expressions for the conductance and the thermoelectric coefficient, recovering exactly the zero temperature limit and showing, at finite temperature, that this quantity is sizeable for strong enough particle-hole asymmetry and in the weak coupling limit.

This paper is organized as follows: In Sec. \ref{sec:model}, we present the system and the physical observable of interest, namely the average current. 
In Sec. \ref{sec:KFT} we use the KFT formalism, after recalling its essential features, to write the exact equations describing the non-equilibrium SYK model, both isolated and coupled to metallic leads, and we analyze the strong coupling analytical solution at equilibrium. In Sec. \ref{sec:current}, we study in detail the current, with considerations on the conductance and the thermoelectric coefficient, and we provide analytical results when possible, as in the tunneling limit or at zero temperature. We give some conclusions in \ref{sec:conclusion}.


\section{Model}
\label{sec:model}

The spinful complex Sachdev-Ye-Kitaev (SYK) model is a zero dimensional quantum mechanical model of $N\gg1$ degenerate orbitals, each of them associated to pairs of creation and annihilation operators, $\hat{d}_{i\sigma}^\dagger$ and $\hat{d}_{i\sigma}$ respectively, related to the $i$-th orbital with $i=1,\dots, N$ and with spin index $\sigma=\uparrow,\downarrow$. The orbitals interact with each other by means of a four-fermion random interaction. The Hamiltonian describing the model is thus 
\begin{equation}
\begin{split}
\label{eq:Hsyk}
    \hat{H}_{dot}=\frac{1}{2}\sum^N_{i,j,k,l=1}\sum_{\sigma,\sigma'}J_{ijkl}\big[\hat{d}^\dagger_{i\sigma'}\hat{d}^\dagger_{j\sigma'}\hat{d}_{j\sigma'}\hat{d}_{l\sigma}
    +\hat{d}^{\dagger}_{l\sigma}\hat{d}^\dagger_{k\sigma'}\hat{d}_{j\sigma'}\hat{d}_{i\sigma}\big]-\mu\sum^N_{j=1}\sum_{\sigma}\hat{d}^\dagger_{j\sigma}\hat{d}_{j\sigma},
\end{split}
\end{equation}
where for simplicity we choose the case in which the interactions are characterized by real and spin-independent coupling constants $J_{ijkl}$ \cite{bib:wang}, which are independent Gaussian random variables with
\begin{equation}
    \overline{J_{ijkl}}=0, \qquad  \overline{J_{ijkl}^2}=\frac{J^2}{(4N)^3},
\end{equation}
and obey the anti-symmetry property
\begin{equation}
    J_{ijkl}=-J_{jikl}=-J_{ijlk}=J_{lkji}.
\end{equation}
Here $J$ is an energy scale regulating the strength of the interactions inside the model. We also explicitly considered the presence of an external field $\mu$, which can be tuned to induce the breaking of particle-hole symmetry in the dot. 
{\color{black} This chemical potential can be experimentally tuned by a gate voltage, which, then, can change the charge, namely the number of fermions, in the system. As we will see in what follows, the conserved charge is related to an infrared parameter, $\theta$, appearing in the conformal limit, which, therefore, controls the particle-hole asymmetry.}
It is easy to see that the theory associated to the SYK Hamiltonian above has a $U(1)$ global symmetry with related conserved charge $\mathcal{Q}$, corresponding to the average fermion number per spin on each orbital 
\begin{equation}
        \mathcal{Q}=\frac{1}{2N}\sum_{i=1}^N\sum_\sigma \langle\hat{d}^\dagger_{i\sigma}\hat{d}_{i\sigma}\rangle, 
\end{equation}
where $\mathcal{Q}\in [0,1]$, 
which can be varied by the field $\mu$.

In our setup, we consider a SYK dot coupled to two metallic leads (labeled as $a=L, R$), which can be modeled as free electron systems. The total system is described by the Hamiltonian $\hat{H}=\hat{H}_{dot}+\hat{H}_{leads}+\hat{H}_T$, where $\hat{H}_{dot}$ is given in (\ref{eq:Hsyk}), while
 \begin{equation}
    \hat{H}_{leads}=\sum_k\sum_\sigma \bigg[\omega_{kL}\hat{c}_{k\sigma L}^\dagger\hat{c}_{k\sigma L} +\omega_{kR}\hat{c}_{k\sigma R}^\dagger\hat{c}_{k\sigma R}\bigg],
\end{equation}
\begin{equation}
    \hat{H}_{T}=\sum_{a=L,R}\sum_k \sum^N_{j=1}\sum_\sigma\bigg[W_{ka,j}\hat{c}^\dagger_{k\sigma a}\hat{d}_{j\sigma}+W^*_{ka,j}\hat{d}_{j\sigma}^\dagger\hat{c}_{k\sigma a}\bigg].
\end{equation}
    Here, the operators $\hat{c}^\dagger_{k\sigma a}$ ($\hat{c}_{k\sigma a}$)  create (annihilate) electrons with momentum $k$, spin $\sigma$, and dispersion relation $\omega_{ka}$ in the corresponding lead $a=L,R$, while $W_{ka,n}$ are tunneling matrix elements between the leads and the SYK dot. In the following, we will consider for simplicity $k$-independent and uniform (i.e. site-independent) tunneling elements, $W_{ka,j}=W_a/\sqrt{N}$ \cite{bib:dellanna}.

Let us apply an external voltage $V$ to the system, which generates a bias in the chemical potentials of the leads, $eV=\mu_L-\mu_R$. We will consider $\mu_L=-\mu_R=eV/2$. 
We can now conveniently 
perform a gauge transformation on the fermionic operators, getting, as a result, a time-dependent hopping parameter 
\begin{equation}
\label{eq:T_gauge}
W_{ka,j}\to W_{ka,j}(t)=W_{ka,j}e^{-i\mu_at}.
\end{equation}
{\color{black} This transformation corresponds to $c_{k\sigma a}\rightarrow c_{k\sigma a} e^{i\mu_a t}$ which is equivalent to performing a unitary transformation of the Hamiltonian, $U=e^{-i t \sum_a(\mu_a N_a)/\hbar}$, where $N_a=\sum_{k\sigma} c^\dagger_{k\sigma a}c_{k\sigma a}$ is the number operator associated to the lead $a$, where $a=R,L$. The Hamiltonian of the leads, including the chemical potentials, $\hat{H}_{leads}=\sum_{a=L,R}\sum_k\sum_\sigma (\omega_{ka}-\mu_a)\hat{c}_{k\sigma a}^\dagger\hat{c}_{k\sigma a}$, is then modified as $U\hat{H}_{leads}U^\dagger-i\hbar (\partial_t U)U^\dagger$. The latter term cancel exactly $\sum_a\mu_aN_a$, namely, by this transformation we can get rid of the chemical potentials, the cost to pay is that the coupling terms between the leads and the dot are time dependent.}\\
We thus define the current operator at the contact $a=L,R$ as
\begin{equation}
\begin{split}
\label{eq:Jtransport}
\hat{J}_a(t)&=-e\frac{d}{dt}\hat{N}_a(t)= -e\frac{d}{dt}\sum_k\sum_\sigma \hat{c}^\dagger_{k\sigma a }(t)\hat{c}_{k\sigma a}(t)
=-ie\sum_\sigma\big[\hat{H}(t),\hat{c}^\dagger_{k\sigma a}(t)\hat{c}_{k\sigma a}(t)\big]\\&=ie\sum_k\sum^N_{j=1}\sum_\sigma \bigg[W_{ka,j}(t)\hat{c}^\dagger_{k\sigma a}(t)\hat{d}_{j\sigma}(t)-W^*_{ka,j}(t)\hat{d}_{j\sigma}^\dagger(t)\hat{c}_{k\sigma a}(t)\bigg],
\end{split}
\end{equation}
where we used the Heisenberg equation of motion and explicitly performed the commutation relation.

\section{SYK Solution}
\label{sec:KFT}
Following Ref. \cite{bib:uguccioni}, we will review the Keldysh field theory formalism, useful also in dealing with quenched disordered systems, like the SYK model. The Keldysh technique 
is a Green's function approach based on the evolution of a system along a closed real time contour $\mathcal{C}$, from $t=-\infty$ to $t=\infty$ and back \cite{bib:keldysh, bib:kamenev}. 
The Keldysh action for the two leads can be constructed in the fermionic coherent state representation, writing it in terms of Grassmann variables $\bar\chi_{k\sigma a}$, $\bar\chi_{k\sigma a}$ corresponding to the operators $\hat{c}^\dagger_{k\sigma a}$, $\hat{c}_{k\sigma a}$, respectively,
\begin{equation}
\begin{split}
    S_{leads}&=\sum_{a=L,R}\sum_k\sum_\sigma\int_\mathcal{C}dt\; \bar\chi_{k\sigma a}(t)(i\partial_t-\omega_{ka})\chi_{k\sigma a}(t)\\&
    =\sum_{a=L,R}\sum_k\sum_\sigma\int^\infty_{-\infty}dt\; \bigg[\bar\chi^+_{k\sigma a}(t)(i\partial_t-\omega_{ka})\chi^+_{k\sigma a}(t)
    -\bar\chi^-_{k\sigma a}(t)(i\partial_t-\omega_{ka})\chi^-_{k\sigma a}(t)\bigg]\\&=\sum_{a=L,R}\sum_k\sum_\sigma\int_\mathcal{-\infty}^\infty dtdt'\; \bar X^T_{k\sigma a}(t)\hat{g}^{-1}_{ka}(t-t')X_{k\sigma a}(t'),
\end{split}
\end{equation}
where in the second line the Keldysh contour is expressed in terms of the $+,-$ branches, while in the third line a Keldysh rotation has been performed, $\chi^{1(2)}=(\chi^+\pm\chi^-)/\sqrt{2}$ and $\bar\chi^{1(2)}=(\bar\chi^+\mp\bar\chi^-)/\sqrt{2}$, and we introduced the vector in Keldysh space $X=\begin{pmatrix}
    \chi^1 &\chi^2
\end{pmatrix}^T$. The inverse Green's function of the isolated leads $\hat{g}^{-1}_{ka}$ are $2\times 2$ matrices in the  Keldysh space, which, in frequency space, reads
\begin{equation}
    \hat{g}^{-1}_{ka}(\omega)=\begin{pmatrix}
        \omega-\omega_{ka}+i0^+ & 2i0^+F_a(\omega) \\ 0 & \omega-\omega_{ka}-i0^+
    \end{pmatrix},
\end{equation}
where 
$F_a(\omega)$ is the distribution function for non-interacting fermions of the lead $a$. 
Under the assumption that the leads are at thermal equilibrium, with temperatures $T_{a}$, the distribution functions are given by 
$F_{a}(\omega)=\tanh[\beta_a\omega/2]\equiv 
1-2f_{a}(\omega)$, where  
$f_a(\omega)=[e^{\beta_a\omega}+1]^{-1}$
is the Fermi function, whit $\beta_a=1/(k_BT_a)$, and $k_B$ the Boltzmann constant. Inverting the matrix $\hat{g}_{ka}^{-1}$ we get the Green's function
\begin{equation}
\label{eq:g_kastruct}
   \hat{g}_{ka}= \begin{pmatrix} g^R_{ka} & g^K_{ka} \\ 0 & g^A_{ka}     
    \end{pmatrix}, 
\end{equation}
\begin{equation}
    g^{R(A)}_{ka}(\omega)=\frac{1}{\omega-\omega_{ka}\pm i0^+},
\end{equation}
\begin{equation}
\label{eq:FDT}
    g_{ka}^K(\omega)=F_a(\omega)[g_{ka}^R(\omega)-g_{ka}^A(\omega)],
\end{equation}
whose components are the retarded ($R$), advanced ($A$) and Keldysh ($K$) Green's functions which, 
being at thermal equilibrium, are linked through the fluctuation-dissipation theorem (FDT), Eq. (\ref{eq:FDT}). Similarly the SYK dot, described by the Grassmann fields $\bar\psi_{j\sigma}$, $\psi_{j\sigma}$ encoding the operators $\hat{d}^\dagger_{j\sigma}$, $\hat{d}_{j\sigma}$, will have the Keldysh action 
\begin{equation}
\begin{split} 
    S_{dot}[\mathbf{J}]&=\int_\mathcal{C} dt\; \bigg[\sum_{j=1}^N\sum_\sigma\Bar{\psi}_{j\sigma}(t)(i\partial_t+\mu)\psi_{j\sigma}(t)\\&-\frac{1}{2}\sum_{i,j,k,l=1}^N\sum_{\sigma,\sigma'}J_{ijkl}\big[\bar\psi_{i\sigma}(t)\bar\psi_{j\sigma'}(t)\psi_{k\sigma'}(t)\psi_{l\sigma}(t)
    +\bar\psi_{l\sigma}(t)\bar\psi_{k\sigma'}(t)\psi_{j\sigma'}(t)\psi_{i\sigma}(t)\big]\bigg], 
\end{split}
\end{equation}
where we did not perform the Keldysh rotation yet, for reasons that will be clear below, and we explicitly considered the dependence from the sets of all couplings $\mathbf{J}=\{J_{ijkl}\}$ in the action.
Finally, the tunneling action reads
\begin{equation}
\begin{split}
    S_{T}&=-\sum_{a=L,R}\sum_k\sum_n\sum_\sigma\int_\mathcal{C} dt\; \bigg[W_{ka,n}(t)\Bar{\chi}_{k\sigma a}(t)\psi_{n\sigma }(t) + W^*_{ka,n}(t)\Bar{\psi}_{n\sigma}(t)\chi_{k\sigma a}(t)\bigg],\\
    &=-\sum_{a=L,R}\sum_k\sum_n\sum_\sigma\int_{-\infty}^\infty dt\; \bigg[W_{ka,n}(t)\bar X^T_{k\sigma a}(t)\Psi_{n\sigma }(t) + W^*_{ka,n}(t)\Bar{\Psi}^T_{n\sigma}(t)X_{k\sigma a}(t)\bigg],
\end{split}
\end{equation} 
where we introduced the Keldysh vector $\Psi=\begin{pmatrix} \psi^1 &\psi^2 \end{pmatrix}^T$ for the SYK spinors. 
By construction, the Keldysh partition function $\mathcal{Z}$ is normalized 
\begin{equation}
    \mathcal{Z}[\mathbf{J}]=\int D[\bar \chi,\chi]D[\bar \psi,\psi]e^{iS[\mathbf{J}]}\equiv 1 
\end{equation}
where $S[\mathbf{J]}=S_{dot}[\mathbf{J}]+S_{leads}+S_{T}$. 
As already, said, $\mathbf{J}$ are random variables, which we assumed Gaussian distributed, therefore can average the Keldysh partition function over the disorder, still having the correct normalization property, namely
\begin{equation}
\begin{split}
    \overline{\mathcal{Z}}=\int D\mathbf{J}\int D[\bar \chi,\chi]D[\bar \psi,\psi]e^{iS[\mathbf{J}]}
    =\int D[\bar \chi,\chi]D[\bar \psi,\psi]e^{iS^{eff}}\equiv1,
\end{split}
\end{equation}
where, in our case, the notation $D\mathbf{J}$ has to be intended as a Gaussian measure, and 
then in the second equality we performed the Gaussian integral over the disorder, getting the effective action $S^{eff}=S_{dot}^{eff}+S_{leads}+S_T$, 
\begin{equation}
\begin{split}
\label{eq:S^eff_dot}
&S^{eff}_{dot}=\sum_{j=1}^{N}\sum_\sigma\int_{\mathcal{C}}dt\;\bar\psi_{j\sigma}(t)(i\partial_t+\mu)\psi_{j\sigma}(t)+\\&+\frac{iJ^2}{4(4N)^3}\int_{\mathcal{C}}dt\int_{\mathcal{C}}dt'\sum^N_{i,j,k,l=1}\bigg[\mathcal{N}_{ijkl}(t,t')+\mathcal{A}_{ijkl}(t,t')\bigg],
\end{split}
\end{equation}
with normal and anomalous terms, respectively
\begin{equation}
    \begin{split}
        \mathcal{N}_{ijkl}&=\sum_{\sigma,\sigma',\rho,\rho'}\big[\bar\psi_{i\sigma}(t)\psi_{i\rho}(t')\big]\big[\bar\psi_{j\sigma'}(t)\psi_{j\rho'}(t')\big]
        \big[\bar\psi_{k\rho'}(t')\psi_{k\sigma'}(t)\big]\big[\bar\psi_{l\rho}(t')\psi_{l\sigma}(t)\big],\\
        \mathcal{A}_{ijkl}&=\sum_{\sigma,\sigma',\rho,\rho'}\big[\bar\psi_{i\sigma}(t)\bar\psi_{i\rho}(t')\big]\big[\bar\psi_{j\sigma'}(t)\bar\psi_{j\rho'}(t')\big]
        \big[\psi_{k\sigma'}(t)\psi_{k\rho'}(t')\big]\big[\psi_{l\sigma}(t)\psi_{l\rho}(t')\big].
    \end{split}
\end{equation}
Notice that in principle one could perform the Keldysh rotation also for the SYK action, which however might be quite cumbersome because it involves the product of four fields (eight fields after averaging over the disorder). For this reason, as we will show below, it is more convenient to derive the equations of the interacting (dressed) Green's function on the whole contour $\mathcal{C}$, and then obtain from it the $R,A, K$ components by means of the so-called Langreth rules \cite{bib:langreth,bib:jauho}, a set of useful identities for practical calculations with functions defined on the closed time contour. 

Let us now consider the isolated SYK-dot (i.e. $W_{a}=0$). In general, there are two equivalent ways of deriving the Dyson equation, namely the equation connecting the dressed $g$ and the bare Green's functions $g_0$: i) the diagrammatic approach and ii) the saddle point condition of an effective action. 
%
We will adopt the second approach. First we need to introduce the bi-local matrix field in the $2\times2$ Nambu space
\begin{equation}
    \tilde{\mathbf{g}}(t,t')=-\frac{i}{N}\sum^N_{j=1}\Phi_j(t)\bar\Phi_j(t')=\begin{pmatrix}
        \tilde g^{11}(t,t') &  \tilde g^{12}(t,t') \\  \tilde g^{21}(t,t') & \tilde g^{22}(t,t')
    \end{pmatrix}, 
\end{equation}
with Nambu vector
$\Phi_i=\begin{pmatrix} \psi_{i\uparrow} & \bar\psi_{i\downarrow} \end{pmatrix}^T$, and thus
\begin{equation}
\begin{split}
    \tilde{g}^{11}(t,t')&=-\frac{i}{N}\sum_{j=1}^N\psi_{j\sigma}(t)\bar\psi_{j\sigma}(t')=-\tilde{g}^{22}(t',t),  \\
    \tilde{g}^{12}(t,t')&=-\frac{i}{N}\sum_{j=1}^N \psi_{j\uparrow}(t)\psi_{j\downarrow}(t')= [\tilde g^{21}(t',t)]^\dagger.
\end{split}
\end{equation}
We can then introduce an identity using functional delta-function
\begin{equation}
\begin{split}
    \mathbf{1}&=\int D\tilde{\mathbf{g}}\,\delta\Big(\tilde{\mathbf{g}}(t,t')+\frac{i}{N}\sum^N_{j=1}\Phi_j(t)\bar\Phi_j(t')\Big)\\&=\int D\tilde{\mathbf{g}}\int D\tilde{\mathbf{\Sigma}}\,\exp\Big\{N\int_\mathcal{C}dt\int_{\mathcal{C}}dt'\,\mbox{tr}\Big[\tilde{\mathbf{\Sigma}}(t',t)
    \Big(\tilde{\mathbf{g}}(t,t')+\frac{i}{N}\sum^N_{j=1}\Phi_j(t)\bar\Phi_j(t')\Big)\Big]\Big\},
\end{split}
\end{equation}
where another independent bi-local matrix field $\tilde{\mathbf{\Sigma}}$ is introduced as a Lagrange multiplier to enforce the delta-function. The averaged Keldysh partition function then takes the form
\begin{equation}
    \overline{\mathcal{Z}}=\int D\tilde{\mathbf{g}}\int D\tilde{\mathbf{\Sigma}}\,e^{iS_{G\Sigma}[\tilde{\mathbf{g}},\tilde{\mathbf{\Sigma}}]},
\end{equation}
where we performed a Gaussian integration over the Grassmann variables $\bar\psi,\psi$ to get the $G\Sigma$ action
\begin{equation}
\begin{split}
    &\frac{S_{G\Sigma}[\tilde{\mathbf{g}},\tilde{\mathbf{\Sigma}}]}{N}=-i\mbox{Tr}\ln\big(\mathbf{1}g^{-1}_0-\tilde{\mathbf{\Sigma}}\big)
    \\&-i\int_\mathcal{C}dt\int_{\mathcal{C}}dt'\,\big[2\tilde{\Sigma}^{11}(t,t')\tilde{g}^{11}(t',t)
    +\tilde{\Sigma}^{12}(t,t')\tilde g^{21}(t',t)
    +\tilde{\Sigma}^{21}(t,t')\tilde g^{12}(t',t)\\&
    -\frac{J^2}{64}\big(\tilde{g}^{11}(t,t')^2\tilde{g}^{11}(t',t)^2+\tilde{g}^{12}(t,t')^2\tilde{g}^{21}(t,t')^2\big)\big].
\end{split}
\end{equation}
The free part of the action is written in terms of the inverse bare propagator on the contour, $g^{-1}_0(t,t')=\delta_{\mathcal{C}}(t,t')(i\partial_t+\mu)$, being $\delta_\mathcal{C}(t,t')$ the contour delta function, defined as $\pm\delta(t-t')$ if the two times are both on the upper (lower) part of the contour, and zero otherwise. From the action above we can obtain the solutions of the stationary point equations $ \delta S_{G\Sigma}/\delta\tilde{\mathbf{\Sigma}}=0$ and $\delta S_{G\Sigma}/\delta\tilde{\mathbf{g}}=0$, namely the solutions $\mathbf{g}$ and $\mathbf{\Sigma}$ satisfying
\begin{equation}
\mathbf{g}(t,t')=\big[\mathbf{1}g^{-1}_0-\mathbf{\Sigma}\big]^{-1}(t,t'),
\end{equation}
and
\begin{equation}
\label{eq:sigmacSYKspin}
\begin{split}
\mathbf{\Sigma}(t,t')=\frac{J^2}{32}\begin{pmatrix}
                g^{11}(t,t')^2g^{11}(t',t) & g^{12}(t,t')^2g^{21}(t',t) \\ g^{21}(t,t')^2g^{12}(t',t) & g^{22}(t,t')^2g^{22}(t',t)
    \end{pmatrix},
\end{split}
\end{equation}
and one can easily check that the anomalous Green's functions and self-energies, $g^{12 (21)}$ and $\Sigma^{12 (21)}$ respectively, admit only the zero solution. Therefore, the only non-trivial equations that survive are the ones for the normal component, and so it is sufficient to study the equations for the $11$ component, which can be rewritten as
\begin{equation}
\begin{split}
\label{eq:sigmacSYKspin2}
  g(t,t')&=\big[g^{-1}_0-\Sigma\big]^{-1}(t,t'), \\ \Sigma(t,t')&=\frac{J^2}{32}g^2(t,t')g(t',t),
\end{split}
\end{equation}
where we dropped the superscripts $11$ to simplify the notation. 
In particular, in the large $N$ limit, the partition function is dominated by the stationary points and therefore these equations become exact. 

The zero solution for the anomalous term implies the absence of superconducting correlations. In order to generate them one has to include some attractive interaction \cite{bib:wang, bib:lantagne} or proximity effect in the presence of superconductors \cite{bib:dellanna}. 
Differently from standard literature \cite{bib:trunin, bib:chowdhury}, in order to derive the Dyson equations, we did not resort to the replica trick, performing the disorder average within the Keldysh formalism. 

In order to evaluate these formal expressions one needs to derive a set of identities which converts products of functions on the contour into products of functions with real-time arguments (i.e. R,A,K components) and contour integrals into standard real-time integrals. This very useful set of identities is known in literature as Langreth rules \cite{bib:jauho,bib:langreth} and allows us to write the Dyson equation in the $2\times 2$ Keldysh space, as in Eq. (\ref{eq:g_kastruct}), in frequency space
\begin{equation}
     \hat{g}(\omega)=\hat{g}_0(\omega)+ \hat{g}_0(\omega)\hat{\Sigma}(\omega)\hat{g}(\omega),
\end{equation}
with retarded and advanced components
\begin{equation}
    g^{R(A)}(\omega)=\frac{1}{\omega+\mu-\Sigma^{R(A)}(\omega)},  
\end{equation}
and Keldysh component
\begin{equation}
    g^K(\omega)=\frac{\Sigma^K(\omega)}{[\omega+\mu-\Re\Sigma^R(\omega)]^2+[\Im\Sigma^R(\omega)]^2}.
\end{equation}
It is important to notice that, from the definition of the Keldysh Green's function \cite{bib:kamenev}, one has the constraint on the conserved charge, namely
\begin{equation}
\label{eq:constrQ}
    g^K(t,t)=-i(1-2\mathcal{Q}).
\end{equation}
Similarly, after some algebra, one can find the $R,A,K$ components for the self-energy
\begin{equation}
\label{eq:sigmacSYKRA1}
\begin{split}
    &\Sigma^{R(A)}(t,t')=-\frac{3J^2}{128}g^{R(A)}(t,t')^2g^{A(R)}(t',t)
    +\frac{J^2}{64}g^{R(A)}(t,t')g^K(t,t')g^K(t',t)\\
    &+\frac{J^2}{128}g^K(t,t')^2g^{A(R)}(t',t)\mp \frac{J^2}{64}g^{R(A)}(t,t')^2g^{K}(t',t)
\end{split}
\end{equation}
and
\begin{equation}
\begin{split}
\label{eq:sigmacSYKK1}
    &\Sigma^K(t,t')=\frac{J^2}{128}g^K(t,t')^2g^K(t',t)
    +\frac{J^2}{128}\bigg[g^R(t,t')^2+g^A(t,t')^2\bigg]g^K(t',t)\\
    &+\frac{J^2}{64}g^K(t,t')\bigg[g^R(t,t')g^A(t',t)+g^A(t,t')g^R(t',t)\bigg]
\end{split}
\end{equation}
Therefore, the above equations describe exactly an isolated spinful complex SYK model in a generic non-equilibrium situation, which can be generally solved numerically.

We can generalize the results obtained above for a SYK-dot coupled to the leads (i.e. $W_{a}\not=0$). Indeed, after performing Gaussian integrations over $\bar\chi,\chi$, one can repeat the same procedure underlined above to easily find 
\begin{equation}
    \begin{split}
    \label{eq:sigmacSYKspin3}
 G_{ij}(t,t')&=\delta_{ij}g_0(t,t')+\int_\mathcal{C} ds\int_\mathcal{C} ds'\, g_0(t,s)\Sigma_{ik}(s,s')G_{kj}(s',t'). \\ \Sigma_{ij}(t,t')&=\frac{J^2}{32}G_{ik}(t,t')G_{kl}(t,t')G_{lj}(t',t)+\sum_{a=L,R}\frac{|W_a|^2}{N}\sum_k g_{ka}(t,t'),
\end{split}
\end{equation}
where the sum over repeated site indices is assumed, and we denoted with $G$ the dressed Green's function of the dot, containing both the SYK interaction and the coupling to the leads. Here $g_{ka}$ is the Green's function for the free lead $a$ defined on the contour, properly obtained by inverting the kernel $g^{-1}_{ka}(t,t')=\delta_\mathcal{C}(t,t')(i\partial_t-\omega_{ka})$. In this case, by applying the Langreth rules, one gets the Dyson equation in Keldysh space 
\begin{equation}
\label{eq:KDeq}
     \hat{G}_{ij}(\omega)=\delta_{ij}\hat{g}_0(\omega)+ \hat{g}_0(\omega)\hat{\Sigma}_{ik}(\omega)\hat{G}_{kj}(\omega),
\end{equation}
and, after some algebra, the $R,A,K$ components for the self-energy
\begin{equation}
\label{eq:sigmacSYKRA2}
\begin{split}
    \Sigma_{ij}^{R(A)}(t,t')&=-\frac{3J^2}{128}G_{ik}^{R(A)}(t,t')G_{kl}^{R(A)}(t,t')G^{A(R)}_{lj}(t',t)
    \\&+\frac{J^2}{64}G^{R(A)}_{ik}(t,t')G_{kl}^K(t,t')G_{lj}^K(t',t)
    +\frac{J^2}{128}G_{ik}^K(t,t')G_{kl}^K(t,t')G_{lj}^{A(R)}(t',t)
    \\&\mp \frac{J^2}{64}G_{ik}^{R(A)}(t,t')G_{kl}^{R(A)}(t,t')G_{lj}^{K}(t',t)
    +\sum_{a=L,R}\frac{|W_a|^2}{N}\sum_k g^{R(A)}_{ka}(t,t'),
\end{split}
\end{equation}
\begin{equation}
\begin{split}
\label{eq:sigmacSYKK2}
   & \Sigma_{ij}^K(t,t')=\frac{J^2}{128}G_{ik}^K(t,t')G_{kl}^K(t,t')G_{lj}^K(t',t)
    \\&\hspace{1cm}+\frac{J^2}{128}\bigg[G_{ik}^R(t,t')G_{kl}^R(t,t')+G_{ik}^A(t,t')G_{kl}^A(t,t')\bigg]G_{lj}^K(t',t)
    \\&\hspace{1cm}+\frac{J^2}{64}G_{ik}^K(t,t')\bigg[G_{kl}^R(t,t')G_{lj}^A(t',t)+G_{kl}^A(t,t')G_{lj}^R(t',t)\bigg]
    +\sum_{a=L,R}\frac{|W_a|^2}{N}\sum_k g^{K}_{ka}(t,t').
\end{split}
\end{equation}
We clearly see that, in this case, the equations for the coupled SYK model are much more complicated to solve with respect to the isolated system, even though a numerical solution is always possible. In the following, we will see that one can always find an analytical solution at strong interaction and large $N$ limit.

\subsection*{Analytical Solution}

The Dyson equations 
written above can be generally solved numerically. However, in the strong interaction limit one can also find an 
analytical solution at equilibrium. In order to find this solution, we introduce the Matsubara Green's function $G^M(\tau)$, where $\tau\in [-\beta/2,\beta/2]$ is the imaginary time, which satisfies a Dyson equation in the form \cite{bib:fetter}
\begin{equation}
\begin{split}
\label{eq:dysonstrongcouplingSYK}
    &G_{ij}^M(\tau,\tau')=\delta_{ij}g^M_0(\tau,\tau')
    +\int_{-\beta/2}^{\beta/2}ds\int_{-\beta/2}^{\beta/2}ds'\,g^M_0(\tau, s)\Sigma_{ik}^M(s,s')G_{kj}^M(s',\tau').
\end{split}
\end{equation}
In the strong interaction limit, $J|\tau-\tau'|\gg1$ and $\beta J\gg 1$, we can drop the inverse free propagator $(g^{M}_0)^{-1}(\tau-\tau')=-\delta(\tau-\tau')(\partial_\tau-\mu)$ in Eq. (\ref{eq:dysonstrongcouplingSYK}) 
{\color{black} (in frequency space $(g^{M}_0)^{-1}\propto (\omega+\mu)$, therefore negligible in the infrared limit, when compared to the self-energy $\propto J^2$)}, getting \cite{bib:SY,bib:kitaev,bib:maldacena,bib:trunin,bib.georges,bib:chowdhury,bib:gu}
\begin{equation}
\label{eq:strongeq}
    \int^{\beta/2}_{-\beta/2} ds\, G_{ik}^M(\tau,s)\Sigma_{kj}^M(s,\tau')\simeq- \delta_{ij}\delta(\tau-\tau'),
\end{equation}
where the SYK self-energy (\ref{eq:sigmacSYKspin3}) for the Matsubara component is given by
\begin{equation}
\label{eq:SigmaMcSYK}
\begin{split}
    \Sigma_{ij}^M(\tau,\tau')&=-\frac{J^2}{32}G_{ik}^M(\tau,\tau')G_{kl}^M(\tau,\tau')G_{lj}^M(\tau',\tau)
    -\sum_{a=L,R}\frac{|W_a|^2}{N}\sum_k g^M_{ka}(\tau,\tau').
\end{split}
\end{equation}
Moreover, for very strong $J$ and in the large $N$ limit, 
we can safely neglect the last term describing the coupling to the leads, producing the same diagonal equations of the isolated SYK model, namely 
\begin{equation}
\begin{split}
\label{eq:strongeasy}
    G_{ij}^M(\tau,\tau')&\simeq \delta_{ij}g^M(\tau,\tau'),\\
    \Sigma_{ij}^M(\tau,\tau')&\simeq\delta_{ij}\Sigma^M(\tau,\tau')=-\frac{J^2}{32}\delta_{ij}g^M(\tau,\tau')^2g^M(\tau',\tau),
\end{split}
\end{equation}
which allow us to rewrite Eq. (\ref{eq:strongeq}) in the following form
\begin{equation}
\label{eq:dysoneqMatsym}
    \frac{J^2}{32}\int^{\beta/2}_{-\beta/2} ds\, g^M(\tau,s)g^M(s,\tau')^2g^M(\tau',s)\simeq\delta(\tau-\tau').
\end{equation}
We also recall that we have the constraint on the conserved charge in Eq. (\ref{eq:constrQ}), which at equilibrium can be written as 
\begin{equation}
\label{eq:GMQ}
    g^M(\tau,\tau+0^+)=\mathcal{Q}.
\end{equation}

In this limit, we can see that the global $U(1)$ symmetry breaks down to a local $U(1)$ symmetry (gauge invariance). Indeed, by neglecting the kinetic term (i.e. the term containing the time derivative) in the action, the $U(1)$ invariance is preserved even in the case where the phases become time dependent. Moreover, one can easily show that the Dyson equation (\ref{eq:dysoneqMatsym}) is invariant under reparametrizations of time $\tau\to f(\tau)$, with $f'(\tau)>0$ (i.e. $f(\tau)$ must preserve the time orientation), as \cite{bib:sachdev,bib:davison}
\begin{equation}
\begin{split}
\label{eq:gauge+reparam}
    g^M(\tau,\tau')&\to g^M[f(\tau),f(\tau')]\frac{h(\tau')}{h(\tau)}f'(\tau)^{1/4}f'(\tau')^{1/4}, \\ \Sigma^M(\tau,\tau')&\to \Sigma^M[f(\tau),f(\tau')]\frac{h(\tau')}{h(\tau)}f'(\tau)^{3/4}f'(\tau')^{3/4},
\end{split}
\end{equation}
where $h(\tau)$ is an arbitrary function representing the emergent local $U(1)$ symmetries. 

To find the proper analytical solution, we start by considering the Fourier component $g^M(\omega_n)$ of the Matsubara Green's function, with discrete frequencies $\omega_n=(2n+1)\pi/\beta$, and we define its analytic continuation to all complex frequencies $z$ via the spectral representation
\begin{equation}
\begin{split}
    \mathcal{G}(z)&=\int_{-\infty}^\infty d\Omega\,\frac{\nu(\Omega)}{z-\omega}, \\ \mathcal{G}(i\omega_n)&\equiv g^M(\omega_n), \qquad \mathcal{G}(\omega\pm i0^+)=g^{R(A)}(\omega),
\end{split}
\end{equation}
where the density of states $\nu(\omega)=-\frac{1}{\pi}\Im \mathcal{G}(\omega+i0^+)$ is always positive for real frequencies. At zero temperature, given the scale invariance implicit in Eq. (\ref{eq:gauge+reparam}), we expect $\mathcal{G}(z)$ to be a power-law of $z$. More precisely, it implies \cite{bib:SY,bib.georges}
\begin{equation}
\label{eq:Gzansatz}
    \mathcal{G}(z)=C\frac{e^{-i(\pi/4+\theta)}}{\sqrt{z}}, \qquad \Im (z)>0,
\end{equation}
{\color{black} 
which is solution of the Dyson equations. The free real parameters $C$ and $\theta$ are then fixed requiring the non-negativity of the corresponding density of states $\nu(\omega)$. This implies the following conditions for $C$ and $\theta$,}
\begin{equation}
    C>0, \qquad 
    -\frac{\pi}{4}\le\theta\le\frac{\pi}{4}.
\end{equation}
The inverse Fourier transform of $\mathcal{G}(i\omega_n)$ yields
\begin{equation}
\label{eq:GMcSYKT=0}
    g^M(\tau)=-C\frac{\mathrm{sgn}(\tau)}{\sqrt{\pi|\tau|}}\sin\bigg(\frac{\pi}{4}+\theta\,\mathrm{sgn}(\tau)\bigg),
\end{equation}
associated to the density of states in the strong coupling limit
\begin{equation}
\label{eq:DOScSYK}
    \nu(\omega)=\frac{C}{\pi\sqrt{|\omega|}}\sin\bigg(\frac{\pi}{4}+\theta\, \mathrm{sgn}(\omega)\bigg).
\end{equation}
The expression above for the DOS clearly shows that $\theta$ is a parameter determining the particle-hole asymmetry associated with the fermionic propagation, forward and backward in time (positive and negative frequencies). For later convenience, it is useful to parametrize this asymmetry in terms of a real parameter $\mathcal{E}$ defined by
\begin{equation}
  e^{2\pi\mathcal{E}}= \frac{\sin(\pi/4+\theta)}{\sin(\pi/4-\theta)}\equiv \tan(\pi/4+\theta),  
\end{equation}
such that $\mathcal{E}=\theta=0$ occurs for particle-hole symmetric case. In particular, it has been shown that the parameter $\mathcal{E}$ is intimately connected to extensive entropy $\mathcal{S}$ (i.e. the $N\to \infty$ limit of the entropy divided by $N$) via the relation \cite{bib:sachdev,bib:davison, bib:gu, bib:chowdhury}
\begin{equation}
\label{eq:entropyE}
    \lim_{T\to0}\frac{d\mathcal{S}}{d\mathcal{Q}}=-\lim_{T\to0}\bigg(\frac{\partial\mu}{\partial T}\bigg)_\mathcal{Q}=2\pi \mathcal{E},
\end{equation}
which is a typical signal that the SYK model corresponds to a non-Fermi liquid (or strange metal), namely it violates the key property of a conventional metals or Fermi liquid, of a vanishing entropy, $\mathcal{S}\rightarrow0$, approaching zero temperature, $T\to0$.

In order to calculate the constant $C$, it is convenient to first find the expression for the Matsubara self-energy $\Sigma^M(\tau)$ from Eq. (\ref{eq:strongeasy}), then compute the Fourier transform, $\Sigma^M(\omega_n)$, and finally perform an analytic continuation to complex frequencies $z$, getting
\begin{equation}
    {\sf\Sigma}(z)=-J^2C^3\frac{e^{i(\pi/4+\theta)}}{32\pi}\cos(2\theta)\sqrt{z},  
\end{equation}
with $\Im(z)>0$, where
\begin{equation}
    {\sf\Sigma}(i\omega_n)=\Sigma^M(\omega_n), \qquad {\sf\Sigma}(\omega\pm i0^+)=\Sigma^{R(A)}(\omega).
\end{equation}
By inserting the expression above, together with the one for $\mathcal{G}(z)$ in Eq. (\ref{eq:Gzansatz}), into the Dyson equation in complex frequencies, which reads
\begin{equation}
    \mathcal{G}(z)=\frac{1}{z+\mu-{\sf\Sigma}(z)},
\end{equation}
we clearly see that, since $\mathcal{G}(z)$ diverges for $z\to0$, one must have the condition $\mu-{\sf\Sigma}(0)=0$, and, by matching the coefficients of the divergent terms, one also gets the following value for the parameter
\begin{equation}
    C=\bigg(\frac{32\pi}{J^2\cos(2\theta)}\bigg)^{\frac{1}{4}}.
\end{equation}

Finally, to determine the particle-hole asymmetry parameter $\theta$, or equivalently $\mathcal{E}$, one can compute the conserved charge $\mathcal{Q}$, i.e. via the constraint in Eq. (\ref{eq:GMQ}) \cite{bib.georges,bib:gu,bib:chowdhury},  
{\color{black}
\begin{equation}
\label{eq:Q}
\mathcal{Q}=\int^\infty_{-\infty}\frac{d\omega}{2\pi i} \,\mathcal{G}(\omega)e^{i\omega0^+}
=iP\int^\infty_{-\infty}\frac{d\omega}{2\pi}\partial_\omega\ln \mathcal{G}(\omega)e^{i\omega0^+}-
iP\int^\infty_{-\infty}\frac{d\omega}{2\pi}\mathcal{G}(\omega)\partial_\omega{\sf\Sigma}(\omega)e^{i\omega0^+}
\end{equation}
where the $P$ symbol means that we take the principal value. This equation can be derived from $\mathcal{G}^{-1}(\omega)=\omega+\mu-{\sf\Sigma}(\omega)$, which implies 
\begin{equation}
\partial_\omega \mathcal{G}^{-1}=1-\partial_\omega {\sf\Sigma}. 
\label{eq:partialG-1}
\end{equation}
Using $\partial_\omega  \mathcal{G}=-\mathcal{G}^2\partial_\omega \mathcal{G}^{-1}$, we, then, can write $\mathcal{G}\partial_\omega \mathcal{G}^{-1}=-\mathcal{G}^{-1}\partial_\omega  \mathcal{G}=-\partial_\omega \ln \mathcal{G}$. From Eq. (\ref{eq:partialG-1}), then we have
\begin{equation}
\mathcal{G}+\partial_\omega \ln \mathcal{G}-\mathcal{G}\,\partial_\omega {\sf\Sigma}=0.
\end{equation}
The first term in the r.h.s. of Eq. (\ref{eq:Q}) is a boundary term which gives $\frac{1}{2}-\frac{\theta}{\pi}$, while the last term, which vanishes for a conventional Fermi liquid, gives $\frac{\sin(2\theta)}{4}$ \cite{bib.georges},} getting
\begin{equation}
\label{eq:genluttinger}
    \mathcal{Q}=\frac{1}{2}-\frac{\theta}{\pi}-\frac{\sin(2\theta)}{4}=\frac{1}{4}\big[3-\tanh(2\pi \mathcal{E})\big]-\frac{1}{\pi}\arctan\big(e^{2\pi\mathcal{E}}\big).
\end{equation}
making $\mathcal{Q}$ a monotonically decreasing function of both $\theta$ and $\mathcal{E}$. We easily notice that the constraints $\theta\in[-\pi/4,\pi/4]$ and $\mathcal{E}\in (-\infty,\infty)$ imply that $\mathcal{Q}\in [0,1]$, as expected. In particular, at the particle-hole symmetric point, $\mathcal{Q}=1/2$ (or equivalently $\mu=0$), one indeed has $\theta=\mathcal{E}=0$. 

The reparametrization invariance (\ref{eq:gauge+reparam}) is very useful to find the exact equilibrium finite-temperature  propagator without solving the corresponding Dyson equation (\ref{eq:dysoneqMatsym}). In particular, we use the map connecting the zero-temperature Euclidean time $\tau \in (-\infty,\infty)$ and the finite temperature Euclidean time $\tau\in[-{\beta/2},{\beta/2}]$, that is $f(\tau)=\tan(\pi \tau/\beta)$, obtaining the 
the following propagator at finite temperature 
\begin{equation}
    g^M(\tau)=-Ch(\tau)\frac{\mathrm{sgn}(\tau)}{\sqrt{\beta|\sin(\pi\tau/\beta)|}}\sin\bigg(\frac{\pi}{4}+\theta\,\mathrm{sgn}(\tau)\bigg),
\end{equation}
where the function $h(\tau)$ is still undetermined apart from a normalization choice $h(0)=1$. Its value can be determined by imposing the anti-periodic condition of the Matsubara Green's function, that is $g^M(\tau+\beta)=-g^M(\tau)$, producing the equation
\begin{equation}
    h(\tau)=h(\tau+\beta)\tan\bigg(\frac{\pi}{4}+\theta\bigg),
\end{equation}
whose solution is 
\begin{equation}
    h(\tau)=\bigg[\tan\bigg(\frac{\pi}{4}+\theta\bigg)\bigg]^{-\frac{\tau}{\beta}}=e^{-2\pi\mathcal{E}\tau/\beta}.
\end{equation}
In the strong coupling limit, therefore, the Matsubara Green's function at finite temperature is given by
\begin{equation}
    g^M(\tau)=-C\frac{e^{-2\pi\mathcal{E}\tau/\beta}}{\sqrt{1+e^{-4\pi\mathcal{E}\,\mathrm{sgn}(\tau)}}}\frac{\mathrm{sgn}(\tau)}{\sqrt{\beta|\sin(\pi\tau/\beta)|}},
\end{equation}
which consistently recovers the zero-temperature correlator (\ref{eq:GMcSYKT=0}) for $\beta\to \infty$. 
By analytic continuation \cite{bib:fetter}, from the Matsubara Green's function we can derive the real time retarded, advanced and Keldysh Green's functions 
\begin{equation}
        g^{R(A)}(t)=\mp iC\frac{\theta(\pm t)e^{\mp i\theta}e^{-2\pi i\mathcal{E}t/\beta}}{\sqrt{\beta\sinh(\pi|t|/\beta)}},  
\end{equation}
\begin{equation}
    g^K(t)= - C\frac{e^{i\theta \, \mathrm{sgn}(t)}e^{-2\pi i\mathcal{E}t/\beta}\,\mathrm{sgn}(t)}{\sqrt{\beta\sinh(\pi|t|/\beta)}}.
\end{equation}

We can finally perform the Fourier transform of these functions above to getting
\begin{equation}
\begin{split}
\label{eq:gRASYK}
    g^{R(A)}(\omega)=\mp iCe^{\mp i\theta}\sqrt{\frac{\beta}{2\pi}}\frac{\Gamma\big(\frac{1}{4}\mp i\frac{\beta\omega}{2\pi }\pm i\mathcal{E}\big)}{\Gamma\big(\frac{3}{4}\mp i\frac{\beta\omega}{2\pi }\pm i\mathcal{E}\big)},
\end{split}
\end{equation}
\begin{equation}
\label{eq:gKSYK}
    g^K(\omega)=-iC\sqrt{\frac{\beta}{2\pi}}F(\omega)\Re\bigg[e^{-i\theta}\frac{\Gamma\big(\frac{1}{4}- i\frac{\beta\omega}{2\pi }+ i\mathcal{E}\big)}{\Gamma\big(\frac{3}{4}- i\frac{\beta\omega}{2\pi }+ i\mathcal{E}\big)}\bigg],
\end{equation}
where $\Gamma(z)$ is the Gamma function, and where we used the FDT 
for the Keldysh Green's function, namely
\begin{equation}
    g^K(\omega)=F(\omega)\big[g^R(\omega)-g^A(\omega)\big],
\end{equation}
with $F(\omega)=\tanh(\beta\omega/2)$, since we assumed thermal equilibrium in the model. Notice that we considered the distribution $F(\omega)$ without potential $\mu$ because we included its effects in the definition of the retarded and advanced Green's functions, via the asymmetry parameters $\theta$ and $\mathcal{E}$, related to the density $\mathcal{Q}$ (and therefore to $\mu$) thanks to the relation in Eq. (\ref{eq:genluttinger}). At zero temperature, $\beta\to \infty$, one finds
\begin{equation}
\label{eq:gRzeroT}
    g^{R(A)}(\omega)=\mp iC\frac{e^{\pm i[\frac{\pi}{4}\mathrm{sgn}(\omega)-\theta]}}{\sqrt{|\omega|}}, 
\end{equation}
\begin{equation}
     g^K(\omega)= -2iC\frac{\mbox{sgn}(\omega)}{\sqrt{|\omega|}}\sin\bigg(\frac{\pi}{4}+\theta\,\mathrm{sgn}(\omega)\bigg),
\end{equation}
where the sign function in the Keldysh Green's function comes from the zero-temperature limit of the fermionic distribution $F(\omega)\xrightarrow{\beta\to\infty} \mbox{sgn}(\omega)$. This zero temperature result is indeed correct, since it exactly reproduces the expression for the DOS $\nu(\omega)=-\frac{1}{\pi}\Im g^{R}(\omega)$ in Eq. (\ref{eq:DOScSYK}).

\section{Currents}
\label{sec:current}
In order to calculate some physical quantities, like the current, let us introduce the auxiliary source fields $A_a$, coupled to 
\begin{equation}
\label{eq:Jfields}
    J_a=ie\sum_{k}\sum_n\sum_\sigma\bigg[W_{ka,n}\bar\chi_{k\sigma a}\psi_{n\sigma}-W^*_{ka,n}\bar\psi_{n\sigma}\chi_{ka\sigma}\bigg],
\end{equation}
corresponding to the operators $\hat{J}_a$ reported in Eq. (\ref{eq:Jtransport}), so to include the following additional term in the action
\begin{equation}
\begin{split}
    S_A&=-\sum_{a=L,R}\int_{\mathcal{C}}dt\; A_{a}(t)J_a(t)
    =-\sum_{a=L,R}\int^{\infty}_{-\infty}dt\; \bigg[A^+_{a}(t)J^+_a(t)-A^-_{a}(t)J^-_a(t)\bigg]\\
    &=-2\sum_{a=L,R}\int^{\infty}_{-\infty}dt\; \bigg[A^{cl}_{a}(t)J^q_a(t)-A^q_{a}(t)J^{cl}_a(t)\bigg],
\end{split}
\end{equation}
where we introduced the classical and the quantum components of the currents $J^{cl(q)}=(J^+\pm J^-)/2$ and of the source fields $A^{cl(q)}=(A^+\pm A^-)/2$. The physical currents, symmetrized over the two branches of the contour, are $J_a^{cl}$, which are coupled to the quantum components of the source fields $A_a^q$, while the quantum components $J^q_a$ are coupled to the classical sources, $A^{cl}_a$, which represent the external physical potentials. 
We can, then, put $A_a^{cl}=0$, since the external voltage $V$ has been already included in the tunneling amplitudes. We can write, therefore, 
\begin{equation}
\begin{split}
    &S_A=-ie\sum_{a=L,R}\sum_k\sum_n\sum_\sigma\int_{-\infty}^\infty dt\; A^q_a(t)\\&
    \times \bigg[W_{ka,n}(t)\Bar{X}^T_{k\sigma a}(t)\hat{\tau}_x\Psi_{n\sigma}(t)-W^*_{ka,n}(t)\Bar{\Psi}^T_{n\sigma}(t)\hat{\tau}_xX_{k\sigma a}(t)\bigg],
\end{split}
\end{equation}
where $\hat{\tau}_x$ is the first Pauli matrix in Keldysh space, namely
\begin{equation}
    \hat{\tau}_x=\begin{pmatrix}
        0 & 1 \\ 1 & 0
    \end{pmatrix}.
\end{equation}
The expectation values of the currents (\ref{eq:Jfields}) can be obtained simply by performing a functional derivative  
\begin{equation}
\label{eq:I_a}
    I_a=\langle J^{cl}_a(t)\rangle=\frac{i}{2}\frac{\delta}{\delta A_a^q(t)}\mathcal{Z}[A_a^q]\bigg|_{A_a^q=0},
\end{equation}
where $\mathcal{Z}[A^q_a]$ is the Keldysh generating functional
\begin{equation}
    \mathcal{Z}[A^q_a]=\int D[\bar{\chi},\chi]D[\bar{\psi},\psi]e^{iS+iS_A},
\end{equation}
such that, for $A_a^q=0$, one recovers the normalized Keldysh partition function, $\mathcal{Z}=\mathcal{Z}[0]=1$.

In our previous work \cite{bib:uguccioni}, we showed in detail that, by differentiating the action over the quantum sources as in Eq. (\ref{eq:I_a}), after performing Gaussian integrations over the free leads, one easily gets the formula for the symmetrized average current in a NESS, also known as Meir-Wingreen formula \cite{bib:meir,bib:wingreen}
\begin{equation}
\begin{split}
\label{eq:meir-wingreenSYK}
    I&=\frac{ie}{2h}\sum^N_{i,j=1}\int_{-\infty}^{\infty}d\omega\; \bigg[\big(\Gamma^L_{ij}(\omega-\mu_L)-\Gamma^R_{ij}(\omega-\mu_R)\big)G_{ji}^K(\omega)
    \\&-\big(F_L(\omega-\mu_L)\Gamma^L_{ij}(\omega-\mu_L)
    -F_R(\omega-\mu_R)\Gamma^R_{ij}(\omega-\mu_R)\big)\big(G^R_{ji}(\omega)-G^A_{ji}(\omega)\big)\bigg],
    \end{split}
\end{equation}
with the energy-dependent couplings defined as
\begin{equation}
\begin{split}
    \Gamma^a_{ij}(\omega)&\equiv2\pi\sum_k W^*_{ka,i}\delta(\omega-\omega_{ka})W_{ka,j}
    =2\pi\frac{|W_a|^2}{N}\nu_a(\omega)\equiv \frac{1}{N}\Gamma^a(\omega),
\end{split}
\end{equation}
and we introduced the density of states (DOS) of the leads $\nu_a(\omega)=\sum_k\delta(\omega-\omega_{ka})$, with $a=L,R$, left and right. The Green's functions $G^p_{ij}(\omega)$, with $p=R,A,K$, are the the retarded, advanced and Keldysh components of the SYK dot coupled to the metallic leads, which can be found via the Keldysh-Dyson equation (\ref{eq:KDeq}), or equivalently
\begin{equation}
\label{eq:KDysonfinal}
    \hat{G}_{ij}(\omega)=\delta_{ij}\hat{g}(\omega)+\sum_{a=L,R}\hat{g}(\omega)\hat{\Sigma}_{a, ik}(\omega)\hat{G}_{kj}(\omega),
\end{equation}
where $\hat{\Sigma}_{a,ij}(\omega)$ is the self-energy contribution coming only from the coupling with the leads, namely  
\begin{equation}
\begin{split}
    \hat{\Sigma}_{a,ij}(\omega)&=\sum_k W^*_{ka,i} \hat{g}_{ka}(\omega-\mu_a)W_{ka,j}
    =\frac{|W_a|^2}{N}\hat{g}_a(\omega-\mu_a)\equiv \frac{1}{  N}\hat{\Sigma}_a(\omega),
\end{split}
\end{equation}
where we defined $\hat{g}_a(\omega)=\sum_k\hat{g}_{ka}(\omega)$, being $\hat{g}_{ka}$ the Green's function for the uncoupled lead $a$ in Keldysh space. On the other hand, $\hat{g}(\omega)$ is the dressed Green's function in Keldysh space for the uncoupled dot with SYK interactions, i.e. with matrix structure
\begin{equation}
    \hat{g}(\omega)=\begin{pmatrix}
            g^R(\omega) & g^K(\omega) \\ 0 & g^A(\omega)
        \end{pmatrix}.
\end{equation}

In what follows we will use the strong interaction limit for the real time SYK Green's functions, obtained in Eqs. (\ref{eq:gRASYK}) and (\ref{eq:gKSYK}), which allow us to derive some analytical results. The price to pay for this choice is that we have to consider only the case where the two electrodes are at the same temperature $T_L=T_R=T$ or, at most, their temperatures differ by a small amount $\Delta T$ (i.e. in linear response). The NESS configuration will be thus generated mainly by the presence of an external voltage $\mu_L-\mu_R=eV$ applied to the electrodes. In this regard, it is still physically correct to use the equilibrium expressions for the Green's functions since, as we already pointed out below, in the dot Hamiltonian (\ref{eq:Hsyk}), the parameter $\mu$ can be seen as an external field tuning the dot out of the particle-hole symmetric point. 
We can, thus, rewrite the Keldysh-Dyson equation (\ref{eq:KDysonfinal}) as
\begin{equation}
    \hat{G}_{ij}(\omega)=\delta_{ij}\hat{g}(\omega)+\frac{1}{N}\hat{g}(\omega)\hat{\Sigma}(\omega)\sum^N_{k=1}\hat{G}_{kj}(\omega),
\end{equation}
where we introduced a comulative self-energy $\hat{\Sigma}=\hat{\Sigma}_L+\hat{\Sigma}_R$ for convenience. To find a solution of this equation, we sum over the site $i=1,\dots,N$, to get
\begin{equation}
    \sum_{i=1}^N\hat{G}_{ij}(\omega)=\hat{g}(\omega)+\hat{g}(\omega)\hat{\Sigma}(\omega)\sum^N_{i=1}\hat{G}_{ij}(\omega),
\end{equation}
with solution
\begin{equation}
    \sum_{i=1}^N\hat{G}_{ij}(\omega)=\big[\hat{1}-\hat{g}(\omega)\hat{\Sigma}(\omega)\big]^{-1}\hat{g}(\omega).
\end{equation}
We now substitute this back into the equation for $\hat{G}_{ij}$, to get the solution
\begin{equation}
    \hat{G}_{ij}(\omega)=\delta_{ij}\hat{g}(\omega)+\frac{1}{N}\hat{g}(\omega)\hat{\Sigma}(\omega)\big[\hat{1}-\hat{g}(\omega)\hat{\Sigma}(\omega)\big]^{-1}\hat{g}(\omega).
\end{equation}

In order to simplify the following calculations for the stationary current, we directly use the wide-band approximation,  for which $g^R_a=-i\pi\nu_a$ for the two electrodes, and we consider a symmetric junction, $\Gamma^L=\Gamma^R=\Gamma$, with $\Gamma^a=2\pi|W_a|^2\nu_a$. In this way, the contribution containing the Keldysh Green's function  $G^K_{ij}$ in the Meir-Wingreen formula (\ref{eq:meir-wingreenSYK}) cancels out, and it is enough to compute the retarded (or advanced) coupled Green's function, whose expression reads
\begin{equation}
\begin{split}
\label{eq:GGamma}
    G_{ij}^{R(A)}(\omega)&=\delta_{ij}g^{R(A)}(\omega)+\frac{1}{N}\mathcal{G}^{R(A)}_\Gamma(\omega),\\\mathcal{G}^{R(A)}_\Gamma(\omega)&\equiv\frac{\Sigma^{R(A)}(\omega)\big[g^{R(A)}(\omega)\big]^2}{1-g^{R(A)}(\omega)\Sigma^{R(A)}(\omega)}=\mp i\Gamma\frac{\big[g^{R(A)}(\omega)\big]^2}{1\pm i\Gamma g^{R(A)}(\omega)}.
\end{split}
\end{equation}

The expectation value of the stationary current through the spinful complex SYK model in the strong interaction limit, for $\mu_L=-\mu_R=eV/2$, can be thus expressed by
\begin{equation}
\begin{split}
\label{eq:I_cSYK}
    &I=-\frac{2e}{h}\frac{\Gamma}{N}\int_{-\infty}^{\infty}d\omega\;\big[f_L(\omega-eV/2)-f_R(\omega+eV/2)\big]\sum_{i,j=1}^N \Im\big[G_{ij}^R(\omega)\big]\\&=-\frac{2e}{h}\Gamma\int_{-\infty}^{\infty}d\omega\;\big[f_L(\omega-eV/2)-f_R(\omega+eV/2)\big]\Im\big[g^R(\omega)+\mathcal{G}^R_\Gamma(\omega)\big].
\end{split}
\end{equation}
The fact that this final expression is independent on the value of $N$ is compatible with the literature, see Ref. \cite{bib:gnezdilov}. 
Now we need to compute the imaginary part of the SYK retarded Green's function $g^R(\omega)$ in Eq. (\ref{eq:gRASYK}). By using the following identity for the Gamma function
\begin{equation}
    \Gamma(1-z)\Gamma(z)=\frac{\pi}{\sin(\pi z)}, \qquad z\not\in \mathbb{Z},
\end{equation}
and choosing $z=\frac{3}{4}-ix$, with $x=\frac{\omega}{2\pi T}-\mathcal{E}$, we can rewrite
\begin{equation}
    g^R(\omega)=-i\frac{Ce^{-i\theta}}{\sqrt{2\pi T}}\bigg|\Gamma\bigg(\frac{1}{4}+ix\bigg)\bigg|^2\sin\bigg(\frac{3\pi}{4}-i\pi x\bigg).
\end{equation}
We then use the identity
   $ \sin(a+ib)=\sin(a)\cosh(b)+ i\cos(a)\sinh(b)$ 
to write the imaginary part of the retarded complex SYK Green's function
\begin{equation}
\begin{split}
\label{eq:imgrSYK}
    &\Im\big[g^R(\omega)\big]=-\frac{C}{2\sqrt{\pi T}}\bigg|\Gamma\bigg(\frac{1}{4}+i\frac{\omega}{2\pi T}-i\mathcal{E}\bigg)\bigg|^2\\&\times\bigg[\cos\theta\cosh\bigg(\frac{\omega}{2T}-\pi\mathcal{E}\bigg)+\sin\theta \sinh\bigg(\frac{\omega}{2T}-\pi\mathcal{E}\bigg)\bigg],
\end{split}
\end{equation}
to be inserted into Eq. (\ref{eq:I_cSYK}). For later convenience, let as also write the real part and the modulus squared, respectively
\begin{equation}
\begin{split}
\label{eq:regrSYk}
    &\Re\big[g^R(\omega)\big]=\frac{C}{2\sqrt{\pi T}}\bigg|\Gamma\bigg(\frac{1}{4}+i\frac{\omega}{2\pi T}-i\mathcal{E}\bigg)\bigg|^2\\&\times\bigg[\cos\theta\sinh\bigg(\frac{\omega}{2T}-\pi\mathcal{E}\bigg)-\sin\theta \cosh\bigg(\frac{\omega}{2T}-\pi\mathcal{E}\bigg)\bigg],
\end{split}
\end{equation}
and
\begin{equation}
\begin{split}
\label{eq:mod2grSYk}
    |g^R(\omega)|^2&=\frac{C^2}{4\pi T}\bigg|\Gamma\bigg(\frac{1}{4}+i\frac{\omega}{2\pi T}-i\mathcal{E}\bigg)\bigg|^4\cosh\bigg(\frac{\omega}{T}-2\pi\mathcal{E}\bigg).
\end{split}
\end{equation}
Finally we need to compute the imaginary part of $\mathcal{G}^R_\Gamma(\omega)$, defined in Eq. (\ref{eq:GGamma}), which reads
\begin{equation}
\begin{split}
\label{eq:ImGGamma}
    \Im\big[\mathcal{G}_\Gamma^R(\omega)\big]&=\frac{\Gamma}{D^R(\omega)}
\bigg(\Im\big[g^R(\omega)\big]^2-\Re\big[g^R(\omega)\big]^2
-\Gamma |g^R(\omega)|^2\Im \big[g^R(\omega)\big]\bigg),
\end{split}
\end{equation}
where 
\begin{equation}
D^R(\omega)=1+\Gamma^2|g^R(\omega)|^2-2\Gamma\Im[g^R(\omega)],
\end{equation}
and where the expressions for $\Im g^R$, $\Re g^R$ and $|g^R|^2$ are reported in Eqs. (\ref{eq:imgrSYK}), (\ref{eq:regrSYk}) and (\ref{eq:mod2grSYk}), respectively.

\subsection{Tunneling Limit}

Let us start the analysis of the current by considering the tunneling limit, namely a regime for which the coupling $\Gamma$ is one of the smallest energy scales in the system. It is easy to see that, in this regime, the contribution in Eq. (\ref{eq:I_cSYK}) containing $\mathcal{G}^R_\Gamma$ can be neglected since is sub-leading with respect to $g^R(\omega)$, and thus the total current can be approximated by
\begin{equation}
    \begin{split}
    \label{eq:I_SYK_tun}
       I&\simeq\frac{e}{h}\frac{\Gamma C}{\sqrt{\pi T}}\int_{-\infty}^{\infty}d\omega\; \big[f_L(\omega-eV/2)-f_R(\omega+eV/2)\big]\bigg|\Gamma\bigg(\frac{1}{4}+i\frac{\omega}{2\pi T}-i\mathcal{E}\bigg)\bigg|^2\\&\times\bigg[\cos\theta\cosh\bigg(\frac{\omega}{2T}-\pi\mathcal{E}\bigg)+\sin\theta \sinh\bigg(\frac{\omega}{2T}-\pi\mathcal{E}\bigg)\bigg].
    \end{split}
\end{equation}
In particular, we will consider two limits for which we can obtain an analytical dependence. 

The first one is the zero-temperature limit, $T\rightarrow 0$, where the Fermi functions become step functions, and we can use the zero-temperature form of the SYK retarded Green's function in Eq. \ref{eq:gRzeroT}, and, thus,
\begin{equation}
    \Im\big[g^R(\omega)\big]=- \frac{C}{\sqrt{|\omega|}}\sin\bigg(\frac{\pi}{4}+\theta\,\mathrm{sgn}(\omega)\bigg),
\end{equation}
getting the following current in the tunneling limit
{\color{black}
\begin{equation}
    \begin{split}
    \label{eq:IuT=0}
       I&\simeq\frac{e}{h}\Gamma C \int_{-eV/2}^{eV/2}
       \sin\bigg(\frac{\pi}{4}+\theta\,\textrm{sgn}(V)\bigg)  \frac{d\omega}{\sqrt{\omega}}
       = \frac{2e^{3/2}}{h}\Gamma C
      \, \cos \theta\, 
       \sqrt{|V|}\,\textrm{sgn}(V)
    \end{split}
\end{equation}
}
where the behavior $I\propto \sqrt{V}$, corresponding to a divergent differential conductance
\begin{equation}
\label{Gtunneling}
    G(V)=\frac{dI}{dV}=\frac{e^{3/2}}{h}\frac{\Gamma C}{\sqrt{|V|}} \,
    {\color{black} \cos\theta}
\end{equation}
represents another signature of a strange metal, with respect to the conventional metallic behavior $I\propto V$. 
{\color{black} The conductance is maximal for $\theta=0$, when particle-hole symmetry is preserved. However, because of the generalized Luttinger theorem \cite{bib.georges,bib:gu}, expressed by Eq. (\ref{eq:genluttinger}), since $\mathcal{Q}\in [0,1]$, $-\pi/4\le \theta\le \pi/4$, then the conductance weakly depends on the particle-hole asymmetry, contrary to what happens for conventional quantum dots.}

The other limit under consideration is the linear response regime, with small voltage $eV$ and temperature bias $\Delta T$. We need to stress again that the condition of a small thermal bias $\Delta T$ corresponds to our limiting case, since it would be incorrect to use the analytical expressions for the Green's functions of the SYK model at thermal equilibrium if one considers an arbitrarily large thermal bias. With this in mind and with the assumptions above, we can write
\begin{equation}
\begin{split}
    f_L(\omega-eV/2)-f_R(\omega+eV/2)&\simeq -\frac{df}{d\omega}(\omega)eV+\frac{df}{dT}(\omega)\Delta T
    =-\frac{df}{d\omega}(\omega)\bigg[eV+\frac{\omega}{T}\Delta T\bigg]
    \\&=\frac{eV+\frac{\omega}{T}\Delta T}{4T\cosh^2[\omega/(2T)]},
\end{split}
\end{equation}
and thus $I\simeq GV+ L\Delta T$, with linear conductance
\begin{equation}
    \begin{split}
    \label{eq:GSYKtunneling}
        G&=\frac{e^2}{h}\frac{C\Gamma}{2\sqrt{\pi}}\frac{\Upsilon(\theta)}{\sqrt{T}},\\
       \Upsilon(\theta)&\equiv \int_{-\infty}^{\infty}d y\;\frac{\big|\Gamma\big(\frac{1}{4}+i\frac{y}{\pi}\big)\big|^2}{\cosh^2(y+\pi\mathcal{E})}\big[\cos\theta\cosh(y)+\sin\theta \sinh(y)\big],
    \end{split}
\end{equation}
where the behavior of $G\propto T^{-\frac{1}{2}}$ represents another hallmark of a non-Fermi liquid. Here, the function $\Upsilon(\theta)=\Upsilon(-\theta)$, is independent on both temperature and bias, and only determines the value of $G$, which becomes an even function of the particle-hole asymmetry parameter $\theta$. It is worth to notice that, in the linear regime, the current is an odd function of $V$ for any value of 
$\theta$. Similarly, one gets the thermoelectric coefficient
\begin{equation}
    \begin{split}
    \label{eq:LSYKtunneling}
        L&=\frac{e}{h}\frac{C\Gamma}{\sqrt{\pi }}\frac{\Xi(\theta)}{\sqrt{T}},\\
        \Xi(\theta)&\equiv \int_{-\infty}^{\infty}d y\;\frac{(y+\pi\mathcal{E})\big|\Gamma\big(\frac{1}{4}+i\frac{y}{\pi}\big)\big|^2}{\cosh^2(y+\pi\mathcal{E})}\big[\cos\theta\cosh(y)+\sin\theta \sinh(y)\big],
    \end{split}
\end{equation}
where we clearly see that $\Xi(\theta)=-\Xi(-\theta)$, and, therefore, the thermoelectric coefficient is an odd function of the parameter $\theta$, which vanishes at the particle-hole symmetric point, $\theta=\mathcal{E}=0$.  This result is compatible with the fact that particle-hole symmetry in a system can strongly influence the thermoelectric effects, with an ideal net cancellation of thermoelectric currents \cite{bib:marchegiani}. This cancellation occurs because for every electron that contributes to the current in one direction due to the temperature gradient, there would be an exactly compensating hole contribution in the opposite direction. Moreover, the odd parity of $L$ physically means that one can control the direction of the thermoelectric current by tuning the external field $\mu$. We can finally compute the Seebeck coefficient $S$, defined in the linear regime, getting the exact value 
\begin{equation}
\label{eq:thermopowertunnexact}
    S\equiv \frac{L}{G}= \frac{2}{e}\frac{\Xi(\theta)}{\Upsilon(\theta)}=\frac{4\pi\mathcal{E}}{3e}=\frac{2}{3e}\ln\bigg[\tan\bigg(\frac{\pi}{4}+\theta\bigg)\bigg],
\end{equation}
which only depends on the particle-hole asymmetry parameter $\mathcal{E}$. This result is perfectly in agreement with the literature \cite{bib:kruchkov}. Since it is known that the parameter $\mathcal{E}$ is strictly connected to the zero temperature extensive entropy $\mathcal{S}$ of the SYK model via the relation (\ref{eq:entropyE}), we get the remarkable result that the thermopower in the SYK model could represent an experimental probe for the evaluation of its residual entropy, since
\begin{equation}
    S=\lim_{T\to0}\frac{2}{3e}\frac{d\mathcal{S}}{d\mathcal{Q}}.
\end{equation}
It is worth to notice that the numerical prefactor $2/3$ is related to our setup, namely a SYK dot weakly connected to two metallic leads. For example, in the case of thermoelectric transport in a lattice made of coupled SYK dots \cite{bib:davison}, the relation between the thermopower and the zero temperature extensive entropy has a slightly different coefficient, that is
    $S=\lim_{T\to0}\frac{1}{e}\frac{d\mathcal{S}}{d\mathcal{Q}}$, 
but still with the same physical meaning, namely that the entropy strongly depends on the particle-hole asymmetry parameter.

It is interesting to notice that the results obtained above for the current in the weak tunneling regime and for the two limits under consideration are perfectly consistent with the scientific literature, in particular with Ref. \cite{bib:can}, where also a numerical study of the current by considering the SYK Green's function far from its analytical limit 
has been performed. It is also a remarkable result showing that we get the same results of Ref. \cite{bib:can}, where the tunneling matrix elements $W_{ka,j}$ are described by random variables, Gaussian distributed, with $\overline{W^2_{ka,j}}$ = $W^2/N$, while we considered $k$-independent and uniform elements, namely $W_{ka,j}=W/\sqrt{N}$ \cite{bib:dellanna}.

\subsection{Linear Response Regime}

We finally consider the case of arbitrary coupling $\Gamma$, but still working within the linear response regime, where $I=GV+L\Delta T$, being $G$ the linear conductance and $L$ the thermoelectric coefficient. More precisely
\begin{equation}
    G=\frac{2e^2}{h}\Gamma\int_{-\infty}^{\infty}d\omega\;\frac{df(\omega)}{d\omega}\,\Im\big[g^R(\omega)+\mathcal{G}^R_\Gamma(\omega)\big],
    \label{G_lr}
\end{equation} 
\begin{equation}
    L=\frac{2e}{h}\frac{\Gamma}{T}\int_{-\infty}^{\infty}d\omega\;\omega\frac{df(\omega)}{d\omega}\,\Im\big[g^R(\omega)+\mathcal{G}_\Gamma^R(\omega)\big],
     \label{L_lr}
\end{equation}
where now we cannot neglect 
$\mathcal{G}^R_\Gamma(\omega)$, defined in Eq. (\ref{eq:GGamma}). 

\begin{figure}
\centering
  \includegraphics[width=0.8\linewidth]{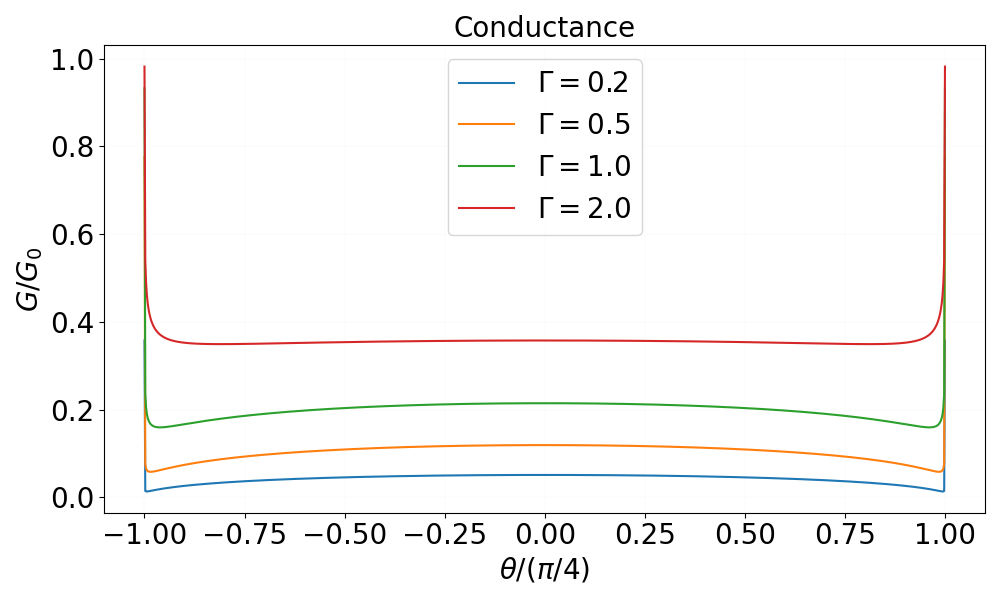}\\
  \includegraphics[width=0.8\linewidth]{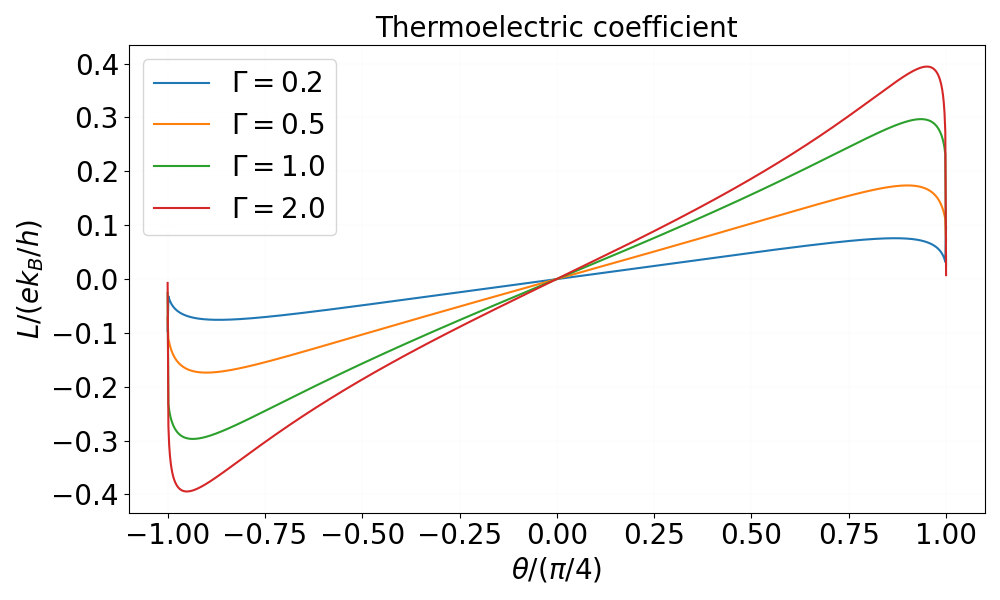}\\
 \includegraphics[width=0.8\linewidth]{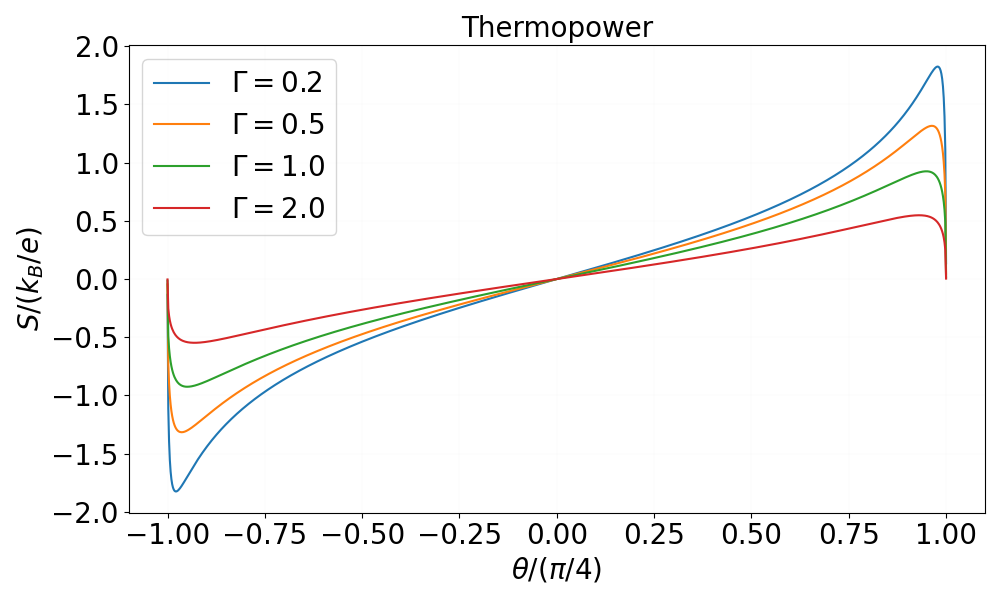}
  \caption{(a) Linear conductance $G$ in units of the conductance quantum $G_0=2e^2/h$, 
 (b) thermoelectric coefficient $L$ in units of {\color{black} $ek_B/h$}, and
 (c) thermopower $S=L/G$ in units of {\color{black} $k_B/e$}, 
  as functions of the particle-hole asymmetry parameter $\theta$, at fixed temperature ($\beta=10$) and for different values of the coupling $\Gamma$. 
  }
  \label{fig:5.1}
\end{figure}

Here below, we present the results obtained through a numerical analysis, where we set the coupling to $J=10^4$ and temperature $k_BT=0.1$ ($\beta=10$). In Figs. \ref{fig:5.1} (a) and \ref{fig:5.1} (b) we show the conductance and thermoelectric coefficient, respectively, as functions of the particle-hole asymmetry parameter $\theta$, for different values of the symmetric coupling $\Gamma$. 
{\color{black} Rather than with multilevel dots, where the thermopower exhibits some oscillations due to the multilevel spectrum \cite{bib:sivan,bib:beenakker92}, our case shares some analogies with a two-fold or $N$ degenerate single level dot, although it has its peculiarities.}  
As expected from the analysis in the tunneling limit, we clearly see that the conductance $G$ is a even function of $\theta$, while the thermoelectric coefficient $L$ is an odd function, in analogy with the transport through a single-level quantum dot, where the particle-hole asymmetry parameter is the dot's energy level with respect to the Fermi energy of the leads \cite{bib:uguccioni}. However, in this case, the conductance seems approximately constant around the particle-hole symmetric point, instead of showing a well-defined peak 
{\color{black} and, contrary to what happens for conventional quantum dots, $G$, for the SYK dot, does not vanish far from particle-hole symmetry, due to the non-vanishing Luttinger-Ward term (the last term in Eq. \ref{eq:Q}).  Actually thermic effects induce narrow spikes in that limit. For small $\Gamma$ we recover the tunneling limit, Eq. (\ref{Gtunneling}). 
Moreover, contrary to the conventional quantum dot, where the thermoelectric coefficient $L$ reaches sizable values for rather small $\Gamma$ ($\Gamma\simeq 2/\beta$) \cite{bib:uguccioni}, here $L$ is large for rather large couplings.}

In Fig. \ref{fig:5.1} (c), we plot the thermopower, defined as the ratio $S=L/G$, as a function of $\theta$ for different values of $\Gamma$. We see that, by decreasing the coupling $\Gamma$, namely by approaching the tunneling limit, the Seebeck coefficient goes to the value reported in Eq. (\ref{eq:thermopowertunnexact}), in agreement with the tunneling limit  shown above. 
{\color{black} Actually, in the latter case, looking at Eqs. (\ref{G_lr}), (\ref{L_lr}), one realizes that $S=L/G\propto \overline\omega/T$. Since $g^R(\omega)$ is peaked at $\omega\sim 2\pi T{\cal E} $ one gets $S\propto \overline\omega/T\sim {\cal E}$.}

\begin{figure}
\centering
  \includegraphics[width=0.8\linewidth]{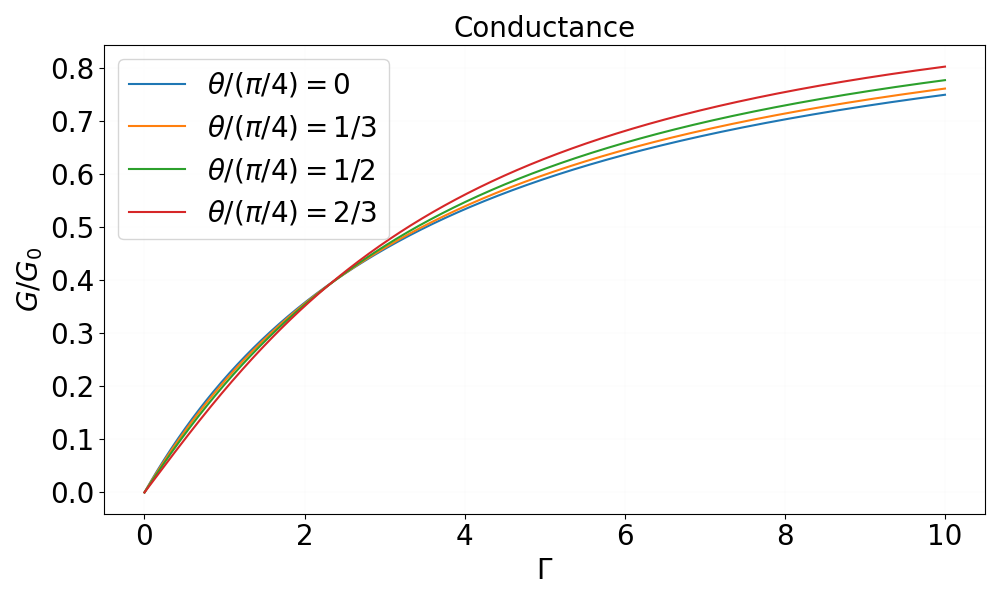}\\
    \includegraphics[width=0.8\linewidth]{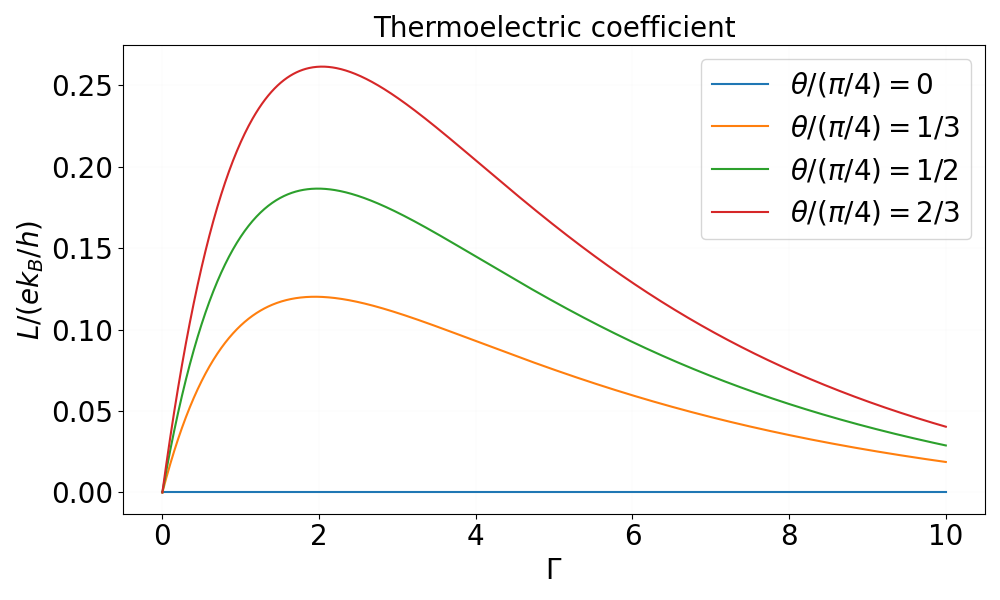}\\
  \includegraphics[width=0.8\linewidth]{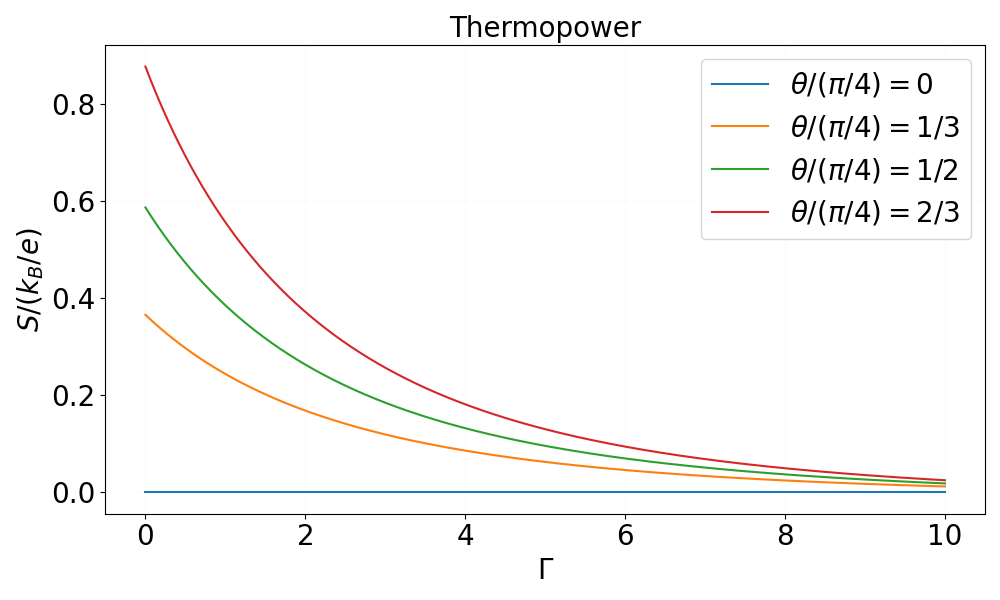}
  \caption{(a) Linear conductance $G$ in units of the conductance quantum $G_0=2e^2/h$, 
(b) thermoelectric coefficient $L$ in units of {\color{black} $ek_B/h$}, and 
(c) thermopower $S=L/G$ in units of {\color{black} $k_B/e$}, 
   as functions of the coupling $\Gamma$, at fixed temperature ($\beta=10$) and for different values of the particle-hole asymmetry parameter $\theta$. 
  }
  \label{fig:5.4}
\end{figure}
In Figs. \ref{fig:5.4} (a), \ref{fig:5.4} (b) and \ref{fig:5.4} (c), we show the conductance, the thermoelectric coefficient, and the thermopower, respectively, as functions of the symmetric coupling between the dot and the leads $\Gamma$, for different values of the asymmetry parameter $\theta$. As expected, for very large coupling (i.e. contact limit), the conductance approaches to the quantum of conductance $G_0=2e^2/h$, 
{\color{black} even far from the particle-hole symmetric point. This is peculiar of the SYK dot. Also the vanishing $G$ for vanishing $\Gamma$ represents the absence of a resonant peak, generally exhibited by conventional quantum dots. On the other hand, the thermoelectric coefficient drops to zero, even for finite $\theta$.} It is interesting to notice that for small and moderate coupling $\Gamma$ the linear conductance is approximately independent on the value of the particle-hole asymmetry parameter $\theta$, 
{\color{black} while the thermoelectric effects are strongly enhanced by the asymmetry, as expected.}

\begin{figure}
\centering
    \includegraphics[width=0.8\linewidth]{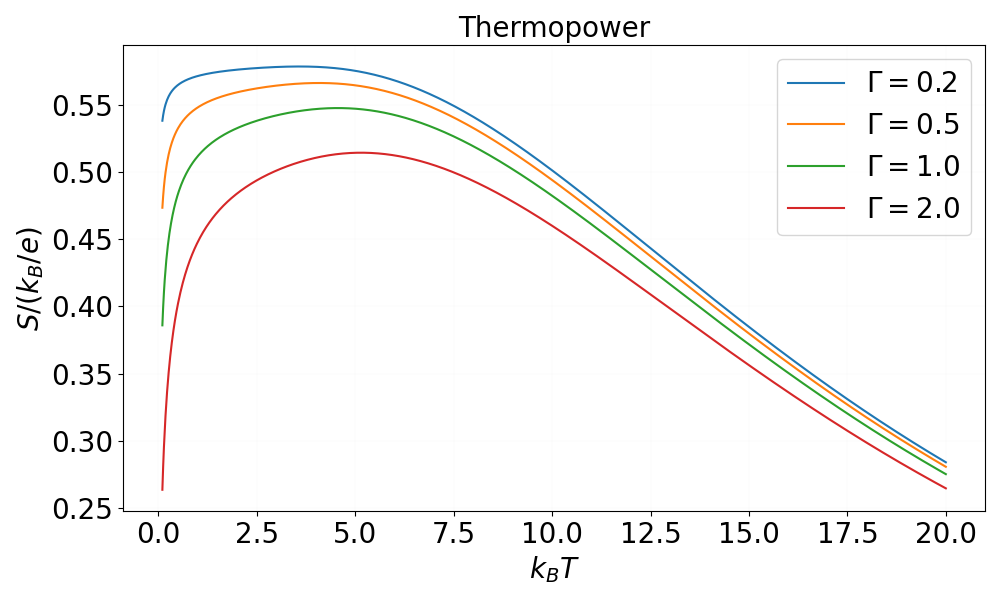}
  \caption{Thermopower $S=L/G$ in units of {\color{black} $k_B/e$} as a function of the temperature, at $\theta=\pi/8$ and for different values of the coupling $\Gamma$.}
  \label{fig:5.7}
\end{figure}

Finally, in Fig. \ref{fig:5.7} we plot the thermopower as a function of the temperature for a fixed value of $\theta$ and different values of $\Gamma$. 
As one can see, 
{\color{black} for $T$ much smaller than $\Gamma$ the curve bends downwards since, while the correlators are smooth, the Fermi distribution becomes narrow and peaked around zero frequency. 
However, for both small $T$ and $\Gamma$ the curve 
becomes flatter and flatter and}
approaches the correct tunneling limit,  
{\color{black} which, in Fig. \ref{fig:5.7}, for $\theta=\pi/8$ is $S\approx 0.58$ (in unit of $k_B/e$), in perfect agreement with the analytic result reported in Eq. (\ref{eq:thermopowertunnexact}). 
}

\section{Conclusions}
\label{sec:conclusion}

In this work, we studied, through the Keldysh field-theory approach, the  transport property of a spinful complex Sachdev-Ye-Kitaev (SYK) dot coupled to metallic leads, forming a N-SYK-N junction. This represents a paradigmatic example of a strongly interacting, disordered quantum system exhibiting non-Fermi-liquid behavior. By KFT we derive the exact Dyson equations without resorting to the traditional replica trick that is often employed in SYK analyses.


The present study extends the scope of KFT to the realm of SYK physics. 
Within this field theory formalism we provide a unified and systematic treatment of electric and thermoelectric transport in a setting that combines strong interactions, quenched disorder, and non-equilibrium driving. Starting from the microscopic Hamiltonian, we derived the full set of self-consistent equations for the Green's functions of the SYK dot. 
In particular, in the tunneling limit we derived exact analytical expressions for the zero-temperature differential conductance, and for both the linear conductance and the thermoelectric coefficient at finite temperature.

Our results show that the SYK island, even when weakly coupled to metallic reservoirs, exhibits clear signatures of its non-Fermi-liquid nature in its transport properties. The electric and thermoelectric responses depend non-trivially on the particle-hole asymmetry parameter, which controls the degree of residual entropy in the system. Remarkably, 
since the link between he thermopower and entropy has been established experimentally for conventional dots \cite{bib.kleeorin}, 
we showed that the thermopower in the SYK model could serve as a direct experimental probe of this residual zero-temperature entropy, thereby linking a measurable transport quantity to one of the most distinctive thermodynamic features of the SYK phase.

Another important outcome of this work is that, contrary to what is typically assumed in the literature, where tunneling matrix elements between the SYK dot and the leads are taken as Gaussian random variables, also constant and uniform tunneling amplitudes lead exactly to the same results. This observation simplifies the theoretical description of the system without loss of generality. 

Moreover, we identified a narrow regime of the coupling to the leads in which the linear conductance becomes approximately independent of the particle-hole asymmetry, while the thermoelectric effects are significantly enhanced. This finding suggests an experimentally favorable configuration for detecting SYK behavior in transport experiments.


In summary, our analysis bridges an important gap between the equilibrium thermodynamics of SYK models and their transport properties under realistic, non-equilibrium conditions. It establishes new analytical connections between entropy and thermoelectric response and provides concrete theoretical guidance for potential experimental realizations in solid-state nanostructures or cold-atom platforms. 

\subsection*{Acknowledgements}
The authors acknowledge financial support from the European Union-Next Generation EU within the "National Center for HPC, Big Data and Quantum Computing" (Project No. CN00000013, CN1 Spoke 10 - Quantum Computing) and from the Project "Frontiere Quantistiche" (Dipartimenti di Eccellenza) of the Italian Ministry for Universities and Research.


\bigskip 

\end{document}